\documentclass{aa}

\usepackage{graphicx}
\graphicspath{{figures/}}
\usepackage{txfonts}
\usepackage{xcolor}
\usepackage{subcaption}
\usepackage{aas_macros}
\usepackage{amsmath}
\usepackage{placeins}

\begin{document}

   \title{Some challenges for the long-term survival of Naiad, Neptune's innermost moon}


   \author{H. Agrusa\inst{1,2}\corrauth{hagrusa@oca.eu}
      \and M. \'Cuk\inst{3}
      \and D. Nesvorn\'y\inst{4}
      \and D. Minton\inst{5}
   }

   \institute{
      Universit\'e C\^ote d'Azur, Observatoire de la C\^ote d'Azur, CNRS, Laboratoire Lagrange, Nice, France
      \and
      Centre national d'\'etudes spatiales (CNES), Paris, France
      \and
      SETI Institute, Mountain View, CA, USA
      \and
      Department of Space Studies, Southwest Research Institute, Boulder, CO, USA
      \and
      Department of Physics and Astronomy, Purdue University, West Lafayette, IN, USA
   }

   \date{Received 18 September 2026 ; accepted }

  \abstract
   {The Naiad-Thalassa 73:69 mean-motion resonance implies these moons have co-existed for $\gtrsim$1~Gyr, raising the question of how they survived to the present day.}
   {We examine three challenges to Naiad's long-term survival: tidal disruption, heliocentric bombardment, and runaway collisional erosion by planetocentric debris.}
   {We constrain Naiad's internal strength requirements from its nominal density and shape, compute present-day impact rates on Neptune's inner moons from heliocentric bombardment, and use $N$-body simulations to model the fate of ejecta produced by non-disruptive impacts.}
   {First, Naiad's nominal density of ${\sim}0.8\text{ g cm}^{-3}$ and elongated shape suggest it cannot be held together by self-gravity alone, implying a cohesive strength of $\gtrsim10$ kPa to avoid tidal disruption, although this constraint is relaxed if Naiad has a higher density of ${\gtrsim}1.3\text{ g cm}^{-3}$. Second, we show that Naiad may have been disrupted in the last 1 Gyr by heliocentric bombardment, although this depends sensitively on the size-frequency distribution of Kuiper Belt objects at small sizes and on Naiad's catastrophic disruption threshold, both of which are poorly constrained. Third, and most notably, we find that even small, non-disruptive impacts can trigger runaway collisional erosion by planetocentric debris on extremely short timescales. Avoiding this ``sesquinary catastrophe'' requires that Naiad has a collisional strength of at least several MPa, which is difficult to reconcile with being a reaccumulated ``rubble pile'', leftover from the capture of Triton and the subsequent cataclysm of Neptune's primordial satellite system.}
   {Together, these results suggest that Naiad may be a physically unusual object among small ring-moons — possibly a largely coherent, monolithic fragment — and that our understanding of Neptune's inner satellite system leaves much to be explained. Future observations and theoretical models of Neptune's inner moons will be critical for resolving these open questions.}


   \maketitle
   \nolinenumbers

\section{Introduction}\label{sec:intro}

The architecture of Neptune's satellite system is thought to originate from the capture of Triton and the resulting cataclysm of Neptune's primordial satellites \citep[e.g.][]{Goldreich1989,Agnor2006}. Gravitational perturbations from Triton would have triggered mutual collisions among Neptune's original regular satellites, forming a debris disk from which the present-day satellites would have accreted \citep{Banfield1992}. This picture is consistent with recent spectroscopic observations of some of Neptune's inner satellites and rings which detected phyllosilicates, a sign of extensive aqueous alteration suggesting that the inner satellites could be the exposed interiors of Neptune's primordial icy satellites \citep{DavisR2026}. In this scenario, the existence of Nereid, Neptune's largest irregular satellite, whose spectrum is inconsistent with being a captured object, can be explained as the sole satellite that survived Triton's capture \citep{Goldreich1989,Belyakov2026a}.

Neptune's five innermost moons -- Naiad, Thalassa, Despina, Galatea, and Larissa -- all orbit within the synchronous orbit, meaning they are slowly migrating inward due to tidal dissipation. Naiad's 4.7$^\circ$ inclination has been thought to be a result of past resonances among other satellites \citep{Banfield1992,ZhangK2008}. However, it was recently discovered that Naiad and Thalassa are currently in a 73:69 mean-motion resonance (MMR), which is the only known fourth-order resonance among planetary satellites \citep{Brozovic2020}. This is an inclination-type resonance, meaning Thalassa's convergent migration is likely responsible for Naiad's inclination.

Under classical tidal theory, Thalassa should migrate inwards faster than Naiad due to its higher mass, despite being slightly further from Neptune \citep[][]{Murray2000}. This means that the convergent migration of Thalassa will pump Naiad's inclination through their resonant interaction, leading to Naiad's present-day ${\sim}5^\circ$ inclination \citep{Cuk2026AJ}. Assuming Naiad was captured into the resonance at a much lower inclination, the orbits of Naiad and Thalassa must have shrunk by ${\sim}20\%$ to explain the present inclination. This implies that Naiad and Thalassa have dynamical ages of order ${\sim}$Gyr, although a higher precision estimate is not possible given the large uncertainties in the past history of Neptune's satellites, their masses, and tidal dissipation rates in Neptune (i.e., its tidal Love number $k_2$ and quality factor $Q$). Another caveat is that Despina, just outside of Thalassa and almost twice its size (therefore ${\sim}8$ times its mass), should migrate faster than Thalassa. \cite{Cuk2026AJ} showed that Despina's convergent migration should disrupt the Naiad-Thalassa MMR on extremely short timescales of only ${\sim}1$ Myr. To alleviate this problem, they speculated that dynamical tides may be active at Neptune, where Thalassa and Despina could be in ``resonance lock'' with some internal oscillation modes of Neptune \citep[e.g.][]{Fuller2016}, forcing them to migrate in lockstep and preventing Despina's convergent migration.

Besides Naiad's peculiar dynamics, it is unusual for several other reasons. At face value, Naiad shouldn't exist; it appears to be within the Roche limit and should have been tidally disrupted by Neptune unless its density and/or cohesive strength is much higher than expected. In addition, bombardment by scattered disk objects (SDOs) and Centaurs may have destroyed it within the last ${\sim}1$ Gyr, which is difficult to reconcile with the old age implied by the Naiad-Thalassa resonance, unless the size-frequency distribution of impactors is shallow at small sizes or Naiad has a high disruption threshold. Finally, and most notably, Naiad's high inclination makes it highly susceptible to runaway collisional erosion by planetocentric debris that could destroy it on short timescales \citep{Cuk2023}. In this paper, we explore these three ``problems'' that Naiad faces, which may shed some light on the origin and evolution of Neptune's satellite system. In Section \ref{sec:naiad_roche} we examine the tidal forces on Naiad to constrain its density and cohesive strength. Then, in Section \ref{sec:naiad_bombardment} we estimate Naiad's collisional lifetime against bombardment from heliocentric impactors. Finally, in Section \ref{sec:naiad_sesquinary} we demonstrate that Naiad should be rapidly destroyed by runaway collisional erosion from planetocentric debris unless it has substantial collisional strength. We briefly discuss the implications of these results in Section \ref{sec:discussion} and conclude in Section \ref{sec:conclusion}.

\section{Naiad's proximity to the Roche limit}\label{sec:naiad_roche}

\begin{figure}
\centering
\begin{subfigure}{0.49\textwidth}
  \centering
  \includegraphics[width=\hsize]{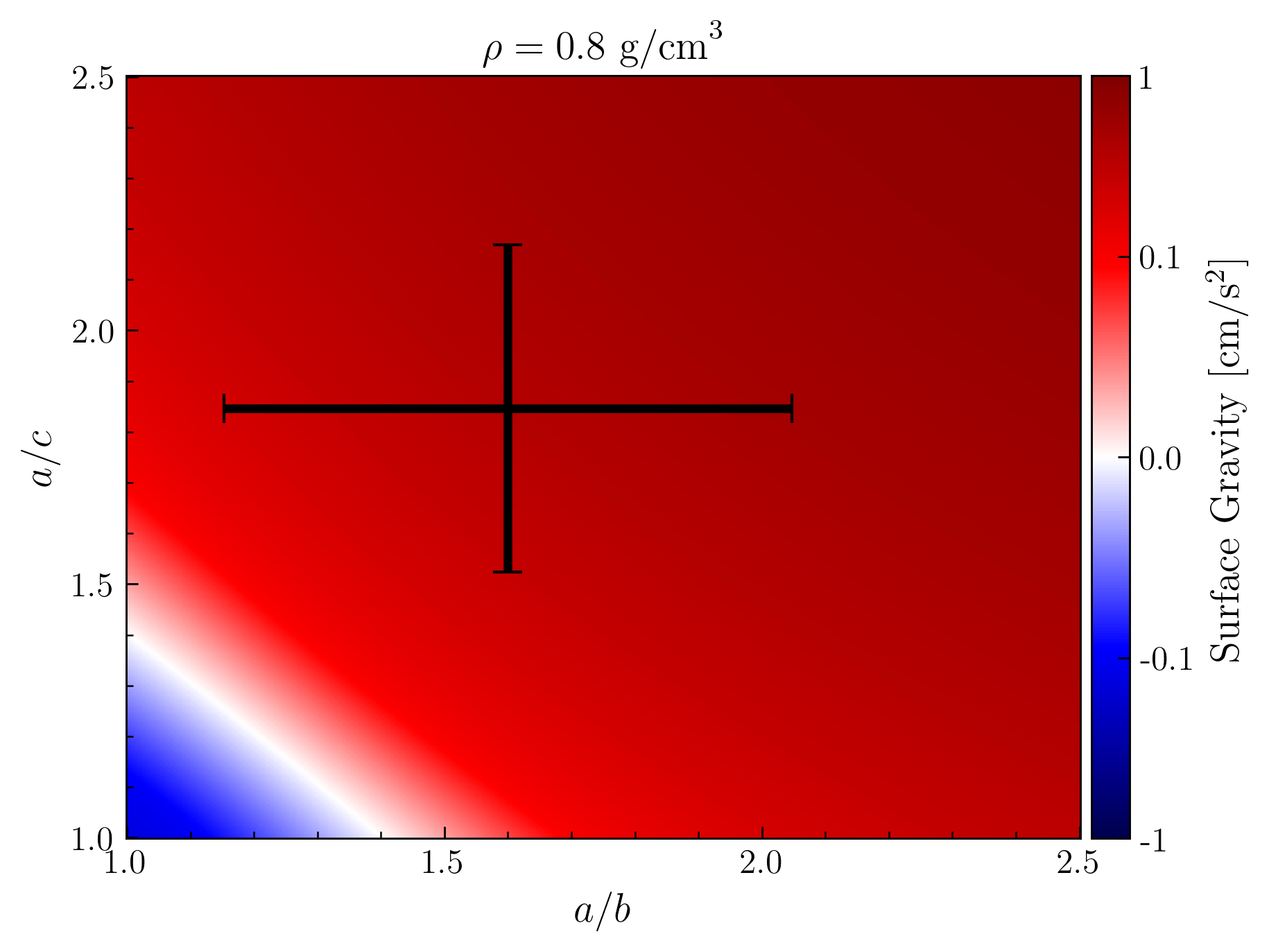}
  \caption{$\rho= 0.8 \text{ g cm}^{-3}$}
\end{subfigure}
\hfill
\begin{subfigure}{0.49\textwidth}
  \centering
  \includegraphics[width=\hsize]{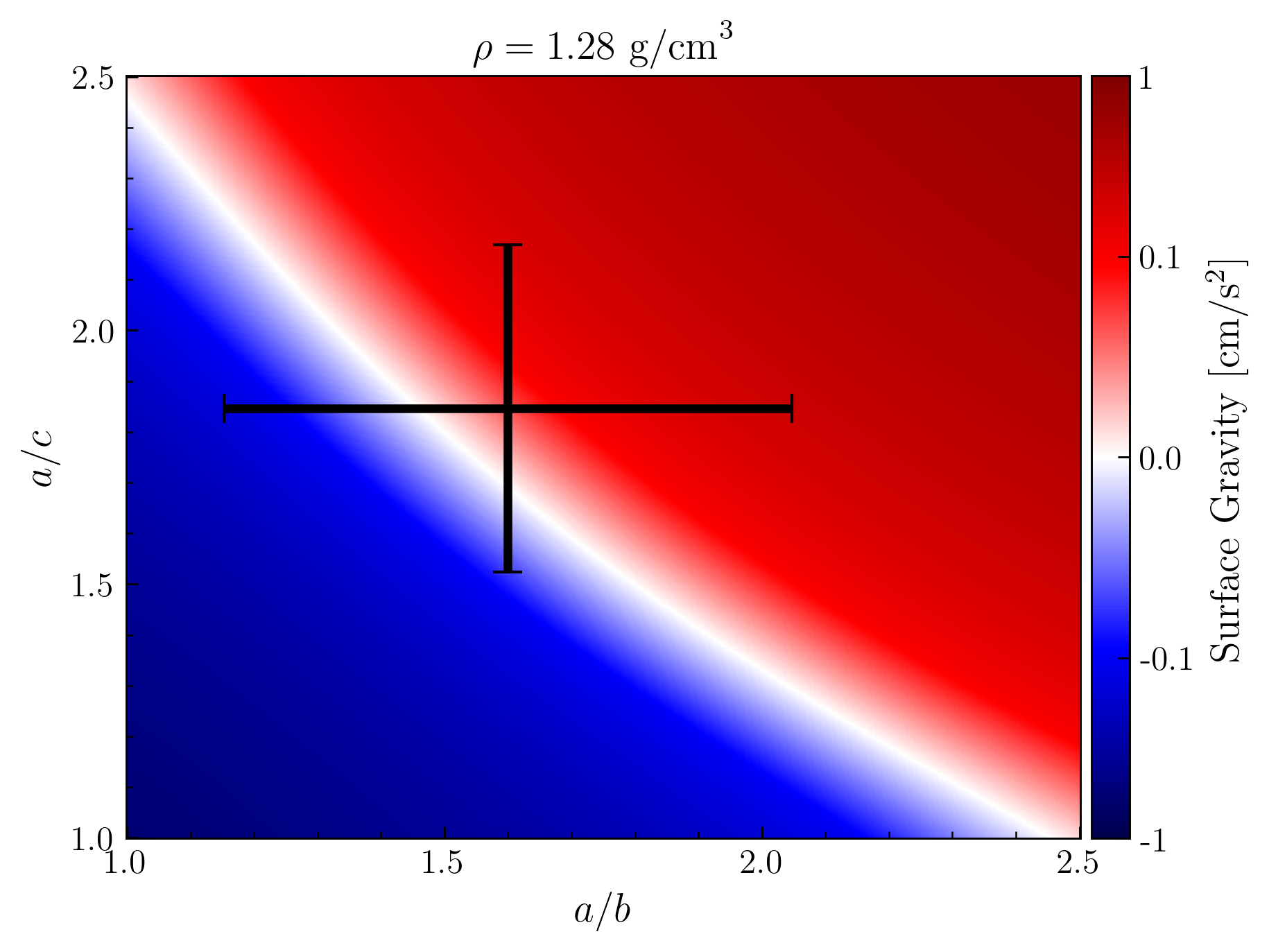}
  \caption{$\rho= 1.28 \text{ g cm}^{-3}$}
\end{subfigure}
\caption{\label{fig:surface_grav} The net surface gravity at the sub-Neptune point on Naiad's surface, depending on its shape, parameterized by $a/b$ and $a/c$. The black point indicates Naiad's nominal shape and its large uncertainty. At Naiad's nominal bulk density of $0.8\text{ g cm}^{-3}$, it is not compatible with being bound by self-gravity alone. At $1.28\text{ g cm}^{-3}$, the upper end of the uncertainty, Naiad's shape must be less elongated than expected to remain gravitationally bound.}
\end{figure}

Based on \textit{Voyager 2} images, the sizes and shapes of Neptune's inner satellites were estimated by \cite{Karkoschka2003}. Although the uncertainties are relatively large, the best-fit ellipsoidal shape of Naiad has semiaxes of $a=48\pm4$ km, $b=30\pm8$ km, $c=26\pm4$ km and an effective radius of ${\sim}33$ km. Recently, \cite{Brozovic2020} were able to estimate the mass of Naiad based on its resonant interaction with Thalassa, where they found $GM_\text{nai}=0.008\pm0.0043 \text{ km}^3 \text{ s}^{-2}$. When combined with the estimate of Naiad's size, this corresponds to a bulk density of $0.8\pm0.48 \text{ g cm}^{-3}$ which is typical for a small icy satellite, albeit with large uncertainties \citep{Thomas2010}. With Naiad's bulk density and shape, we can then compute the net gravitational acceleration on Naiad's surface at the sub-Neptune point to estimate how close Naiad is to tidal disruption. We can write the net acceleration as $a_\text{net}=a_\text{grav}+a_\text{tides}+a_\text{rot}$, where $a_\text{grav}$ is the self-gravitational acceleration due to Naiad, $a_\text{tides}$ is the tidal acceleration due to Neptune, and $a_\text{rot}$ is the centrifugal acceleration due to Naiad's tidally-locked rotation state. Rather than treat it as a sphere, we approximate Naiad's self-gravity due to its ellipsoidal shape using MacCullagh's formula \citep{Murray2000}. Then, the acceleration is written as,

\begin{multline}
  a_\text{net} = -\frac{GM_\text{nai}}{a^2} - \frac{3G}{2a^4}\big(B+C-2A\big) + \frac{GM_\text{nep}}{(d-a)^2}
   -\frac{GM_\text{nep}}{d^2}  \\+ \frac{G(M_\text{nep}+M_\text{nai})}{d^3}a,
\end{multline} 
where $M_\text{nep}$ and $M_\text{nai}$ are the respective masses of Neptune and Naiad, $a$, $b$, and $c$ are the semi-axis lengths of Naiad corresponding to the moments of inertia $A$, $B$, and $C$, and $d$ is the distance between Neptune and Naiad. The first two terms are Naiad's self-gravity, the next two are the tidal acceleration, and the last term is the centrifugal acceleration. More details can be found in \cite{Agrusa2026a}. In Fig.\ \ref{fig:surface_grav}, we plot the net acceleration at the sub-Neptune point as a function of Naiad's axis ratios $a/b$ and $a/c$ for different bulk densities. In these figures, Naiad's ellipsoidal shape and approximate uncertainties are given by the black marker and errorbar. At the nominal bulk density of 0.8 g cm$^{-3}$, loose material at Naiad's sub-Neptune point is unbound for any plausible shape, placing Naiad within its classical Roche limit. At the upper end of the uncertainty on Naiad's density, at 1.28 g cm$^{-3}$, it seems Naiad may be able to hold itself together if its shape is slightly less elongated than the nominal value.

\begin{figure}
\centering
\includegraphics[width=0.5\textwidth]{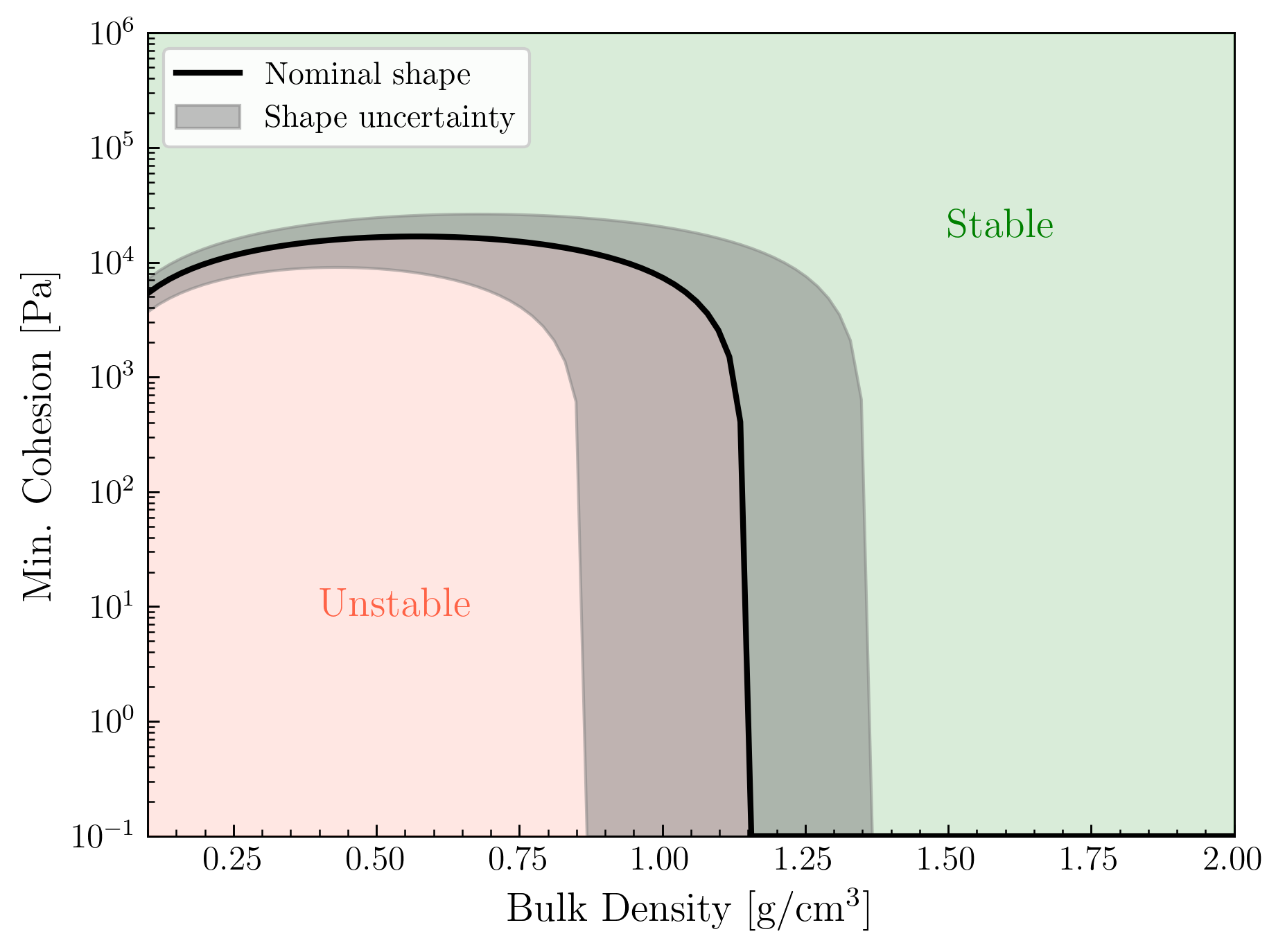}
\caption{\label{fig:min_cohesion} The minimum required cohesive strength for Naiad as a function of its bulk density, assuming a friction angle of $35^\circ$. The black line corresponds to Naiad's nominal ellipsoidal shape with axis ratios $a/b=1.6, a/c=1.85$, while the gray shaded region is an envelope encompassing the uncertainty in Naiad's shape. }
\end{figure}

One should interpret these plots with some caution, as there are still significant uncertainties in Naiad's mass, bulk density, and shape, but they do demonstrate that Naiad may be at or even within the Roche limit, unlike any other satellite in the Solar System for which we have a measured shape and mass. This suggests that Naiad may require some internal strength to exist today \citep{Hedman2015}. If Naiad can be treated as a granular material, as might be expected if it reaccumulated from collisional debris originating from the destruction of a regular satellite system, we can estimate its minimum required cohesive strength using the Drucker-Prager yield criterion \citep{Holsapple2006,Holsapple2008}. This calculation is shown in Fig.\ \ref{fig:min_cohesion}, where we plot the minimum required cohesive strength to avoid failure as a function of Naiad's bulk density, assuming a friction angle of $35^\circ$, which is typical for granular materials \citep{Beakawi2018}. Although highly dependent on its shape and model assumptions, we find that Naiad requires a bulk cohesive strength of $\gtrsim10$ kPa at its nominal shape. It is notable that this cohesive strength is several orders of magnitude higher than the inferred strengths of small bodies recently visited by spacecraft, although Naiad has a very different size and origin than these objects \citep[e.g.,][]{Groussin2015,Groussin2019,Arakawa2020,Walsh2022,Perry2022}. A low, ${\sim}0.8$ g cm$^{-3}$ bulk density for Naiad would be consistent with it being an icy satellite, but this would also require it to have substantial cohesive strength which is difficult to explain. If Naiad has a slightly higher bulk density, of ${\sim}1.3$ g cm$^{-3}$ or larger, then no cohesive strength is required and it can remain intact with self-gravity alone. This possibility is intriguing, given recent observations of Larissa, Galatea, and Proteus that show no evidence of water ice and strong 2.72 $\mu$m absorption bands indicating the presence of phyllosilicates \citep{DavisR2026}. If Naiad has a similar composition and is not an icy satellite, then it would likely have a high enough density to alleviate the need for substantial cohesive strength. Together, these results suggest that Naiad must have either a higher density, a less elongated shape, substantial cohesive strength, or some combination of the three.

\section{Naiad's collisional lifetime}\label{sec:naiad_bombardment}

The apparent old age of the Naiad-Thalassa 73:69 resonance implies that both of these objects have survived collisional bombardment by Centaurs and SDOs for ${\sim}$Gyr timescales. Owing to Naiad's high orbital speed and gravitational focusing by Neptune, heliocentric impactors collide at extremely high speeds of ${\sim}20$ km/s \citep{Zahnle2003}, meaning that relatively small impactors have the potential to destroy Naiad. 

Early works demonstrated that the collisional lifetimes of Naiad and Thalassa are less than the age of the Solar System, implying they have undergone multiple disruptions and should be rubble piles today \citep{Colwell1992}. We revisit the impact rate on Neptune and Naiad's collisional lifetime an updated model for the outer Solar System \citep{Nesvorny2023}. Briefly, \cite{Nesvorny2023} developed a dynamical model of long-period comets, ecliptic comets (ECs), and Centaurs by tracking the orbital evolution of test particles from the trans-Neptunian scattered disk through the inner solar system using $N$-body simulations spanning 4.5 Gyr. The model was calibrated against observations of Centaurs from the Outer Solar System Origins Survey (OSSOS) and against the known population of active ECs \citep{Bannister2018}. Impact probabilities on each moon were then computed accounting for the moons' orbital eccentricities, inclinations, physical sizes, and gravitational focusing by Neptune. The resulting impact probabilities are expressed relative to Jupiter, which can then be converted to an absolute impact rate by adopting a reference rate for Jupiter based on the calibrated EC population. From this model, we obtain an impact rate on Naiad for objects larger than 10 km in size, $P^\text{naiad}_\text{imp}({>}10\text{ km})=1.2\times10^{-11}\text{ yr}^{-1}$. The full list of impact rates and speeds are listed in Table \ref{tab:moons}.

\begin{table}
\caption{Impact probabilities and orbital properties of select Neptune satellites\label{tab:moons}}
\centering
\begin{tabular}{lcccc}
\hline\hline
Moon & $P_{\rm imp}(>10\text{ km})$ & $V_{\rm imp}$ & $a$ & $R$ \\
     &                              & (km s$^{-1}$) & ($\times10^5$ km) & (km) \\
\hline
Naiad      & $3.3\times10^{-7}$ & $20.7$ & $0.482$    & $33.0$   \\
Thalassa   & $4.1\times10^{-7}$ & $20.4$ & $0.501$    & $40.0$   \\
Despina    & $6.3\times10^{-6}$ & $20.3$ & $0.525$    & $74.0$   \\
Galatea    & $1.4\times10^{-6}$ & $19.7$ & $0.62$    & $79.0$   \\
Larissa    & $2.6\times10^{-6}$ & $16.7$ & $0.735$    & $96.0$   \\
Hippocamp  & $2.6\times10^{-7}$ & $15.5$ & $1.053$   & $35.0$   \\
Proteus    & $7.0\times10^{-6}$ & $14.6$ & $1.176$   & $208.0$  \\
Triton     & $9.9\times10^{-5}$ & $7.7$  & $3.548$   & $1352.6$ \\
Nereid     & $2.5\times10^{-7}$ & $3.3$  & $55.138$  & $170.0$  \\
Halimede   & $6.8\times10^{-9}$ & $3.0$  & $166.871$ & $31.0$   \\
\hline
\end{tabular}
\tablefoot{Column definitions: $P_{\rm imp}$ is the impact probability relative to Jupiter, $V_{\rm imp}$ is the mean impact velocity, $a$ is the orbital semi-major axis, and $R$ is the moon radius. The impact rate on Jupiter is $P_\text{imp}({>}10\text{ km})=3.5\times10^{-5} \text{yr}^{-1}$.}
\end{table}

Using $P^\text{naiad}_\text{imp}({>10}\text{ km})$, we can extrapolate to smaller sizes given an assumed impactor size-frequency distribution (SFD). In order to estimate collision rates for small impactors, we first assume the SFD of the Kuiper Belt is well-represented by the Jupiter Trojan SFD, which should be a good proxy if the Trojans are captured KBOs that were scattered inwards during the giant planet instability \citep{Tsiganis2005,Morbidelli2005}. Between ${\sim}2$ and ${\sim}10$ km, the Trojan cumulative SFD is well-characterized by a power-law of the form $N(d)\propto d^{-2.1}$, although the exact size cutoffs and power law indices vary between studies \citep[e.g.][]{Grav2011,Wong2015,Yoshida2017,Uehata2022}. Below this size, the SFD cannot be constrained with telescopic surveys, but impact craters on Charon and Arrokoth from the New Horizons mission can provide some constraints. 

The crater counts on these two bodies suggest a break to a shallow slope in the crater SFD  below ${\sim}13$ km (corresponding to impactors below ${\sim}2$ km), although there is some uncertainty and debate over how shallow this slope is \citep{Singer2019,Spencer2020,Morbidelli2021,Robbins2021}. These works often report the power-law index of the differential SFD of craters, which must be converted to an index for the cumulative impactor SFD. If the differential crater power-law index has the value $q$, then the cumulative crater power-law index is $q+1$. Then, we adopt a crater-to-projectile scaling law of the form $D\propto d^{0.8}$, although there is considerable uncertainty in this scaling depending on the unknown material properties of Charon and Arrokoth \citep{Housen2011,Morbidelli2021}. Given a differential crater SFD with power-law index of $q$, then the cumulative impactor SFD would have a power-law index of ${\sim}0.8(q+1)$. For example, \cite{Robbins2021} report that small craters on Charon's Vulcan Planitia have a differential SFD with index $q=-1.7\pm0.2$, which corresponds to a cumulative impactor SFD index of $q_\text{imp}\approx-0.56\pm0.16$, ignoring any additional uncertainty introduced by the crater-to-projectile scaling law. Putting it all together, the impact rate on Naiad can be estimated as:

\begin{equation}
  P(d) = \begin{cases}
  P(d>10\text{ km})\bigg(\frac{d}{10\text{ km}}\bigg)^{-2.1} & \text{if } 2 \text{ km} \leq d \leq 10\text{ km}  \\
  P(d>10\text{ km})\bigg(\frac{2\text{ km}}{10\text{ km}}\bigg)^{-2.1}\bigg(\frac{d}{2\text{ km }}\bigg)^{q_\text{imp}} & \text{if }  d \leq 2\text{ km}  \\
\end{cases}
\end{equation}

Next, we need to estimate the impactor size capable of disrupting Naiad. The impactor diameter ($D_\text{imp}$) required to destroy a target of size $D_\text{targ}$ can be estimated as $D_\text{imp}=(2Q^*_D/v_\text{imp}^2)^{1/3}D_\text{targ}$, where $v_\text{imp}$ is the impact speed and $Q^*_D$ is the catastrophic disruption threshold. $Q^*_D$ is the specific impact energy required to disperse half the mass involved in a collision, leaving the other half in the largest remnant \citep{Benz1999}. The catastrophic disruption threshold has a complicated dependence on the physical and material properties of both the target and projectile and the impact conditions. $Q^*_D$ can be measured at small scales with laboratory experiments, or estimated with numerical simulations, or inferred on a population level from collisional evolution models \citep[e.g.,][]{Holsapple2002,Jutzi2015,Bottke2005}. To compute some representative examples for Naiad's $Q^*_D$, we use the size-dependent $Q^*_D$ scaling law of \cite{Bottke2020} which was modified from \cite{Benz1999}:
\begin{align}
  Q^*_D(R) &= aR^\alpha+bR^\beta \\
  a &= \frac{Q^*_{D_{LAB}}}{R^\alpha_{LAB}}\frac{1}{1-\frac{\alpha}{\beta}(\frac{R_{LAB}}{R_{min}})^{\beta-\alpha}} \\
  b &= -\frac{\alpha}{\beta}aR_{min}^{\alpha-\beta},
\end{align}
where $R$ is the radius of the body, $\alpha$ and $\beta$ determine the respective slope of the $Q^*_D$ curve in the strength and gravity regime, $R_{min}$ defines the location where $Q^*_D$ is a minimum. Finally, $Q^*_{D_{LAB}}$ and $R_{LAB}$ correspond to laboratory experiments where $Q^*_D$ was measured in order to anchor the scaling law at smaller sizes. \cite{Bottke2005} used this size-dependent scaling law to reproduce the observed size-frequency distribution of the main belt. According to their best-fit scaling law for the main belt, and if Naiad is ``asteroid-like'' in terms of impact strength, an object with 33 km radius would have $Q^*_D{\sim}6\times10^8$ erg/g. Similarly, \cite{Bottke2023} derived an empirical scaling law for the Kuiper Belt constrained by the size-frequency distribution of Trojan asteroids and the cratering record on the giant planet satellites. According to this scaling, a ``KBO-like'' Naiad would have a disruption threshold of $Q^*_D\approx3\times10^7$ erg/g. At typical collision speeds for Naiad, this corresponds to 4 and 1.5 km-sized impactors being able to disrupt Naiad if it is ``asteroid-like'' or ``KBO-like'', respectively. 

Naiad's close-in orbit means that its effective $Q^*_D$ could be further reduced due to Neptune's tidal potential. \cite{Agrusa2025b} found that ``$Q^*_{TD}$'', the tidally-dependent disruption threshold, scales as $Q^*_{TD}\sim Q^*_D(1-\delta_\text{Roche}^3/\delta^3)$, where $Q^*_D$ is the disruption threshold without the influence of tides, and $\delta$ and $\delta_\text{Roche}$ are the dimensionless distances of the Roche limit and the satellite's orbit, respectively. They are expressed in physical units as $\delta=\frac{d}{R_P}(\frac{\rho}{\rho_P})^{(1/3)}$, where $d$ is the satellite's orbital distance, $R_P$ is the central body's radius, and $\rho$ and $\rho_P$ are the respective bulk densities of the satellite and central body. This expression for $Q^*_{TD}$ scales with the inverse cube of the distance and goes to zero at the Roche limit as one might expect. Of course, Naiad's disruption threshold is unknown, as is its Roche limit, so it is impossible to actually estimate $Q^*_{TD}$. At Naiad's nominal bulk density, it has a normalized orbital distance of $\delta{\approx}1.54$, and for generic rocks and soil, the closest a spherical satellite can orbit is about $\delta_\text{Roche}{\approx}1.5$ \citep{Holsapple2006}. Using these numbers as a rough guide, it means Neptune's tides could reduce Naiad's effective disruption threshold by another factor of ${\sim}10$, although this effect could be much larger or much smaller. In this case, a ``KBO-like'' Naiad could be disrupted by a ${\sim}0.6$ km-sized impactor rather than a ${\sim}1.5$ km impactor.

\begin{figure}
\centering
\includegraphics[width=0.5\textwidth]{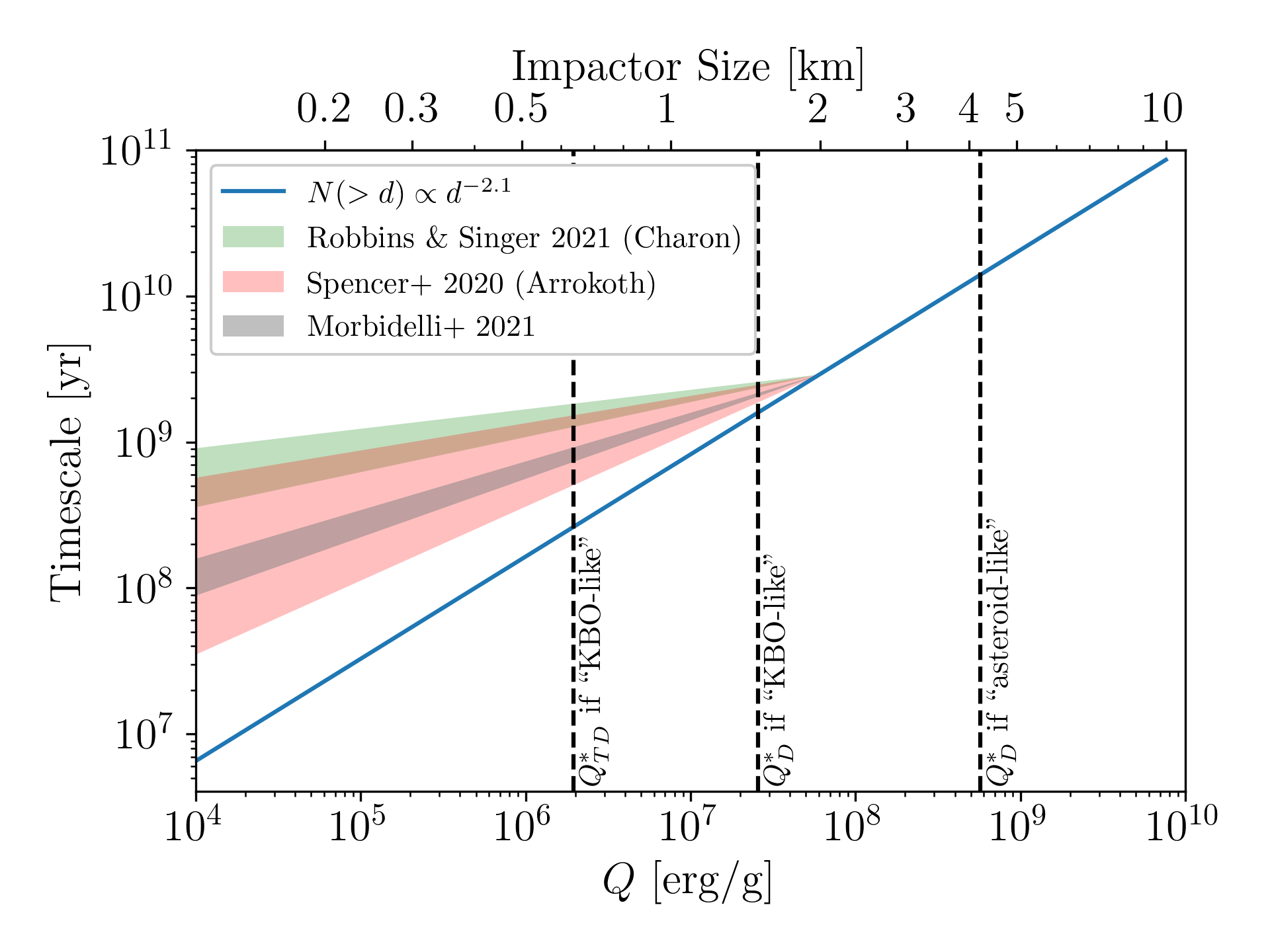}
\caption{\label{fig:naiad_lifetime} The timescale for impacts on Naiad of a given specific energy or size, extrapolated from the impact rate for objects $>10$ km, based on possible (and highly uncertain) size-frequency distributions of impactors at smaller sizes.}
\end{figure}

In Fig.\ \ref{fig:naiad_lifetime} we plot the collision timescale (the inverse of the collision rate) as a function of impactor size under different assumptions for the impactor size-frequency distribution at small sizes. We show the full uncertainty ranges for the power-law indices reported by \cite{Spencer2020} (Arrokoth craters), \cite{Robbins2021} (Charon craters) and \cite{Morbidelli2021} (combined), although no additional uncertainty is added when converting from their reported crater SFD to an impactor SFD. We also plot several representative values for Naiad's $Q^*_D$ and the corresponding impactor size \citep{Bottke2005,Bottke2023,Agrusa2025b}.

So long as Naiad's collisional strength is sufficiently high or the SFD of small impactors is sufficiently shallow, then it can survive for ${\sim}$Gyr and its dynamical age and collisional lifetime can be reconciled. The uncertainties in both the impactor population at small sizes and in Naiad's $Q^*_D$ are huge, and this figure should be interpreted bearing that in mind. The bigger challenge to Naiad's collisional survival may lie in what happens after small, non-disruptive impacts that should be relatively frequent, which we explore in the following section.

\section{Sesquinary impacts} \label{sec:naiad_sesquinary}

Due to Naiad's high inclination and orbital velocity, it is susceptible to a ``sesquinary catastrophe'', whereby impact ejecta can leave Naiad's surface at a low relative speed, undergo differential nodal and apsidal precession due to Neptune's oblateness, and then reimpact at high speed, leading to runaway collisional erosion \citep{Cuk2023}. Naiad's short ${\sim}7$ h orbit period means that this process would operate on very short timescales. Although large disruptive collisions may be rare over Naiad and Thalassa's lifetimes, both objects should be regularly bombarded by small impactors which can trigger runaway collisional erosion. One obvious way to avoid sesquinary catastrophe is by having sufficient collisional strength, which we investigate here.

$N$-body simulations have proven effective at capturing the collisional runaway process of sesquinary catastrophe in the case of Deimos around Mars and we apply similar methods here \citep{Anand2026}. Here, we use the REBOUND $N$-body code to model the short-term dynamical and collisional evolution of Naiad \citep{rebound}. Ejecta particles are treated as test particles and REBOUND was modified to skip the collision search between test particles, which enables the simulations to run with $\mathcal{O}(N)$ complexity. The effects of Neptune's $J_2$ and $J_4$ were included with ReboundX and the simulations were integrated using the TRACE integrator \citep{reboundx,reboundtrace}. The simulations included only Neptune, Naiad, Thalassa, and Despina, along with test particles. Higher-order effects such as perturbations from other moons and solar tides were neglected. The orbits of Naiad, Thalassa, and Despina are initialized using the latest SPICE kernels provided by NAIF\footnote{\url{https://naif.jpl.nasa.gov/naif/index.html}} \citep{SpiceyPy2020,NAIF}. We adopt values of $J_2=3409.1\times10^{-6}$, $J_4=-33.4\times10^{-6}$ \citep{Brozovic2020,JacobsonR2009} and the masses and radii for Neptune and its inner three moons are given in Table \ref{tab:body_params}. The simulations are integrated for $10^6$ Naiad orbital periods, or ${\sim}800$ years.

Each simulation begins with a single 10 m diameter impactor striking Naiad's leading face at 20 km/s, which is a typical speed for heliocentric impactors. We then follow the evolution and production of additional ejecta from sesquinary impacts. Any collisions that occur below the escape speed of the target result in a perfect inelastic merger, otherwise new ejecta are initialized according to the point-source crater scaling laws of \cite{Housen2011}. We adopt these scaling relations (referred to hereafter as HH11) as a simple way to connect Naiad's strength to collision outcomes. These relations provide the crater radius ($R$) in either the strength- or gravity-dominated regimes as a function of the impactor radius ($a$), velocity ($U$), density ($\delta$), mass ($m$), target density ($\rho$), impact strength ($Y$), and surface gravity ($g$). In the strength and gravity regimes, the respective crater radii are,

\begin{align}
  R_\mathrm{grav} &= H_1
    \left(\frac{\rho}{\delta}\right)^{\!\frac{2+\mu-6\nu}{3(2+\mu)}}
    \!\left(\frac{g a}{U^2}\right)^{\!-\frac{\mu}{2+\mu}}
    \!\left(\frac{m_i}{\rho}\right)^{\!1/3}, \label{eq:Rgrav}\\[4pt]
  R_\mathrm{str}  &= H_2
    \left(\frac{\rho}{\delta}\right)^{\!\frac{1-3\nu}{3}}
    \!\left(\frac{Y}{\rho U^2}\right)^{\!-\frac{\mu}{2}}
    \!\left(\frac{m_i}{\rho}\right)^{\!1/3}, \label{eq:Rstr}
\end{align}

where $\mu$ and $\nu$ are dimensionless parameters which relate how the impactor velocity and density affect the resulting crater, respectively. $H_1$ and $H_2$ are additional dimensionless normalization constants that can be determined from laboratory experiments. By equating these two expressions and solving for $Y$, we can determine the critical strength $Y_\text{crit}$ defining the transition between the gravity and strength regimes:

\begin{equation}
  Y_\mathrm{crit} = \bigg(\frac{H_2}{H_1}\bigg)^{2/\mu}\bigg(\frac{\delta}{\rho}\bigg)^\frac{2\nu}{2+\mu}\bigg(\frac{ga}{U^2}\bigg)^\frac{2}{2+\mu}\rho U^2. \label{eq:Ycrit}
\end{equation}

For a given impact, if the target strength is below $Y_\text{crit}$, we consider the impact to be in the gravity regime and use Eq.\ \ref{eq:Rgrav} to compute the resulting crater radius, otherwise, we use Eq.\ \ref{eq:Rstr}. Given the crater radius, the ejecta mass and velocity distributions can be calculated as a function of distance $x$ from the impact point:
  
\begin{align}
   M(x) &=  \frac{3 k\rho}{4\pi\delta}\left[\left(\frac{x}{a}\right)^3 - n_1^3\right] m_i, \label{eq:Mx}\\
   v(x) &= C_1 \left[\frac{x}{a}\left(\frac{\rho}{\delta}\right)^\nu\right]^{-1/\mu}
         \left(1 - \frac{x}{n_2 R}\right)^p U, \label{eq:vx}
\end{align}

where $n_1$, $n_2$, $C_1$, $k$, and $p$ are additional constants. These relations are only valid over the interval $n_1a\leq x \leq n_2R$. Inside of $n_1a$ and outside of $n_2R$, the power-law behavior is expected to break down due to the effects of strength and/or gravity. These relations for $M(x)$ and $v(x)$ can then be used parametrically to determine the cumulative ejecta mass-velocity distribution $M(>v)$. For simplicity, we adopt the best-fit scaling parameters derived from SPH simulations of the DART impact on the asteroid Dimorphos \citep{Raducan2024b,Cheng2024}, which are listed in Table \ref{tab:ejecta_params}. Although Dimorphos and Neptune's inner moons likely have very different physical and material properties, we note that these scaling parameters are all on the order of unity and do not vary substantially for different target materials \citep{Raducan2019}.

\begin{table}
\caption{Point-source scaling parameters for impact ejecta taken from \cite{Cheng2024}\label{tab:ejecta_params}}
\centering
\begin{tabular}{ccccccccc}
\hline\hline
$\nu$ & $\mu$ & $n_1$ & $n_2$ & $k$ & $H_1$ & $H_2$ & $C_1$ & $p$ \\
\hline
0.4 & 0.45 & 1.2 & 1.3 & 0.42 & 0.396 & 0.34 & 0.47 & 0.3 \\
\hline
\end{tabular}
\end{table}

\begin{table}
\caption{Body parameters used in $N$-body simulations. The masses are taken from \cite{JacobsonR2009} and \cite{Brozovic2020}, while the radii come from \cite{Karkoschka2003}.\label{tab:body_params}}
\centering
\begin{tabular}{lcc}
\hline\hline
Body & $GM$ (km$^3$ s$^{-2}$) & Mean Radius (km) \\
\hline
Neptune  & 6{,}835{,}100 & 24{,}764 \\
Naiad    & 0.0080        & 33       \\
Thalassa & 0.0236        & 41       \\
Despina  & 0.1179        & 75       \\
\hline
\end{tabular}
\end{table}
To keep these simulations computationally tractable and avoid introducing additional unconstrained physics, we make several simplifying assumptions. All ejecta particles are assigned a fixed size of 10 m with bulk density equal to that of the target. Their velocities are drawn from the cumulative mass-velocity distribution $M(v)$ given by Eqs. \ref{eq:Mx} and \ref{eq:vx}, treated as a cumulative distribution function and sampled by inverse transform sampling. Because the ejecta cone geometry is not easily described by analytic models like HH11, each particle is launched in a direction drawn randomly and uniformly from the hemisphere defined by the outward surface normal at the impact point. Since the ejecta particles do not interact with each other, they are all initialized at the same surface location. Once initial conditions are set, the target's mass and velocity are adjusted to conserve mass and momentum. This simple collision model does not perfectly conserve angular momentum, but because the ejecta mass is small relative to Naiad and the simulations span short timescales, this effect is negligible.
  
The most significant simplification is the fixed particle size, which ignores the size-frequency distribution of real ejecta. If the SFD is steep, mutual collisions could grind ejecta to dust before it reimpacts Naiad, which would then be removed by Poynting-Robertson (PR) drag. A more sophisticated collisional evolution model should be explored in future work, but this simple model likely captures the essential physics. Even if collisional grinding were significant, it would produce a dusty ring in the Laplace plane that Naiad would continue to pass through at high velocity, driving further erosion. For collisional grinding to halt the sesquinary catastrophe, PR drag would need to remove dust faster than the collision timescale with Naiad, which is unlikely given the short collisional timescales demonstrated in our simulations.

\begin{figure*}
\centering
\begin{subfigure}{0.33\textwidth}
  \centering
  \includegraphics[width=\hsize]{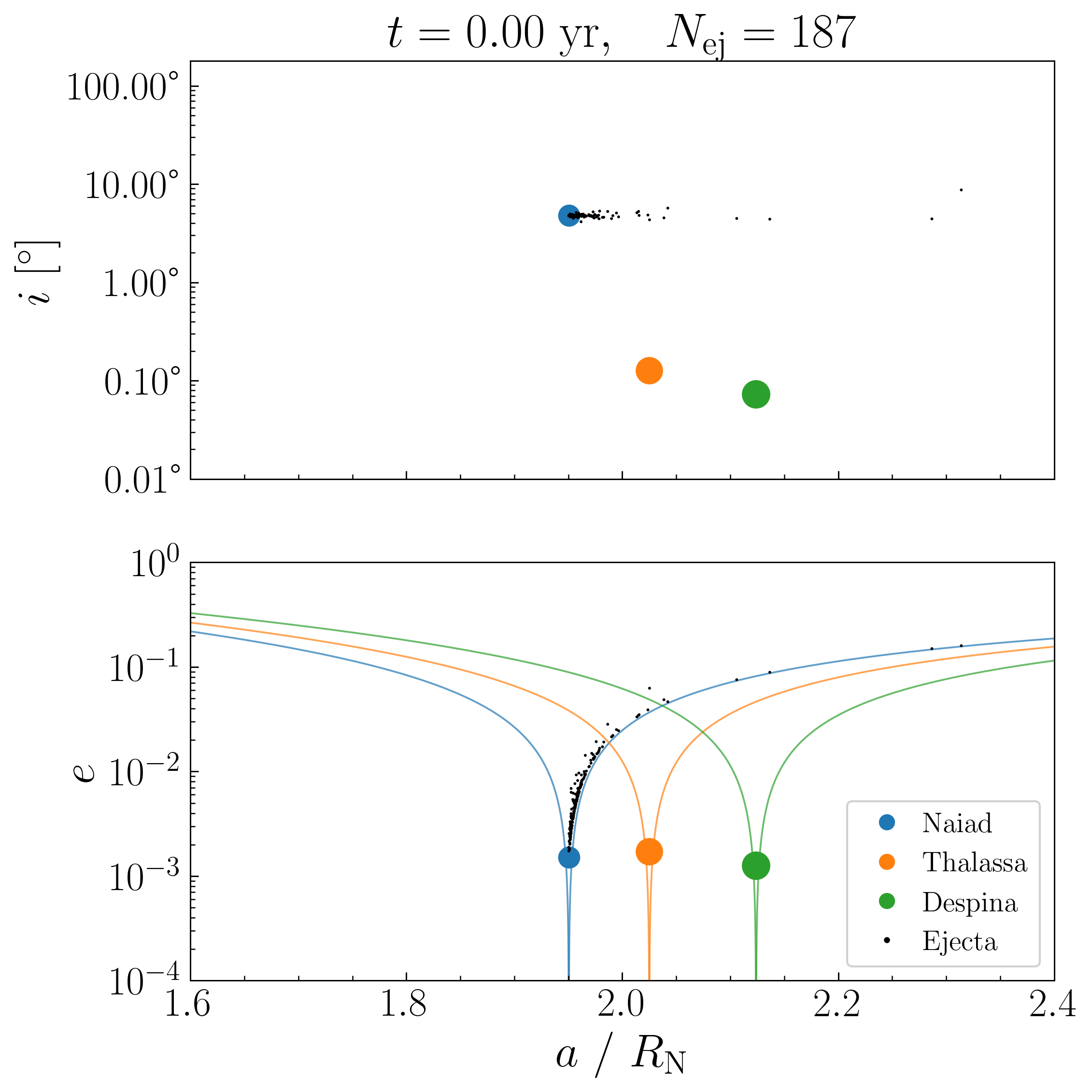}
  \caption{$t = 0$}
\end{subfigure}
\begin{subfigure}{0.33\textwidth}
  \centering
  \includegraphics[width=\hsize]{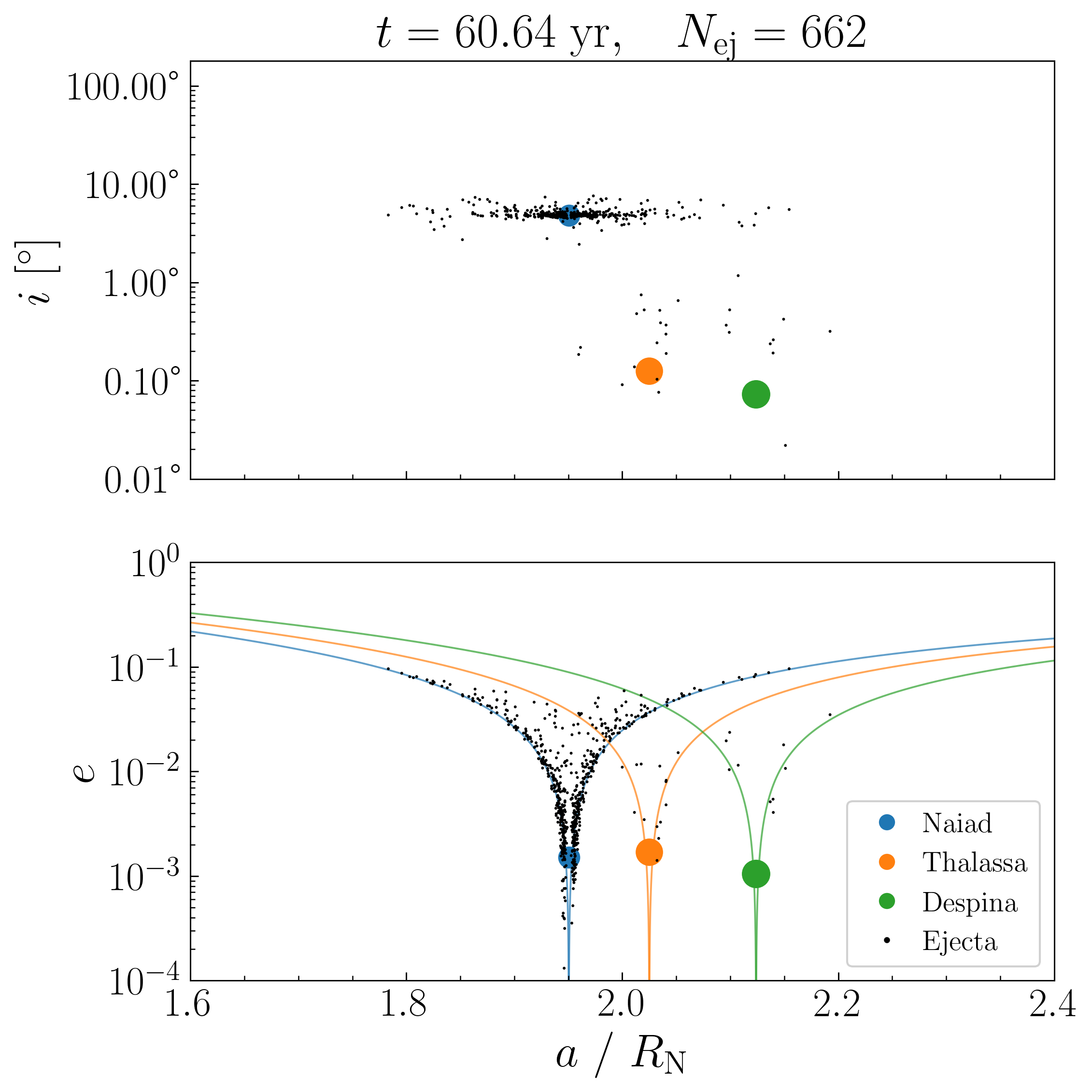}
  \caption{$t \sim 60$ yr}
\end{subfigure}
\\[6pt]
\begin{subfigure}{0.33\textwidth}
  \centering
  \includegraphics[width=\hsize]{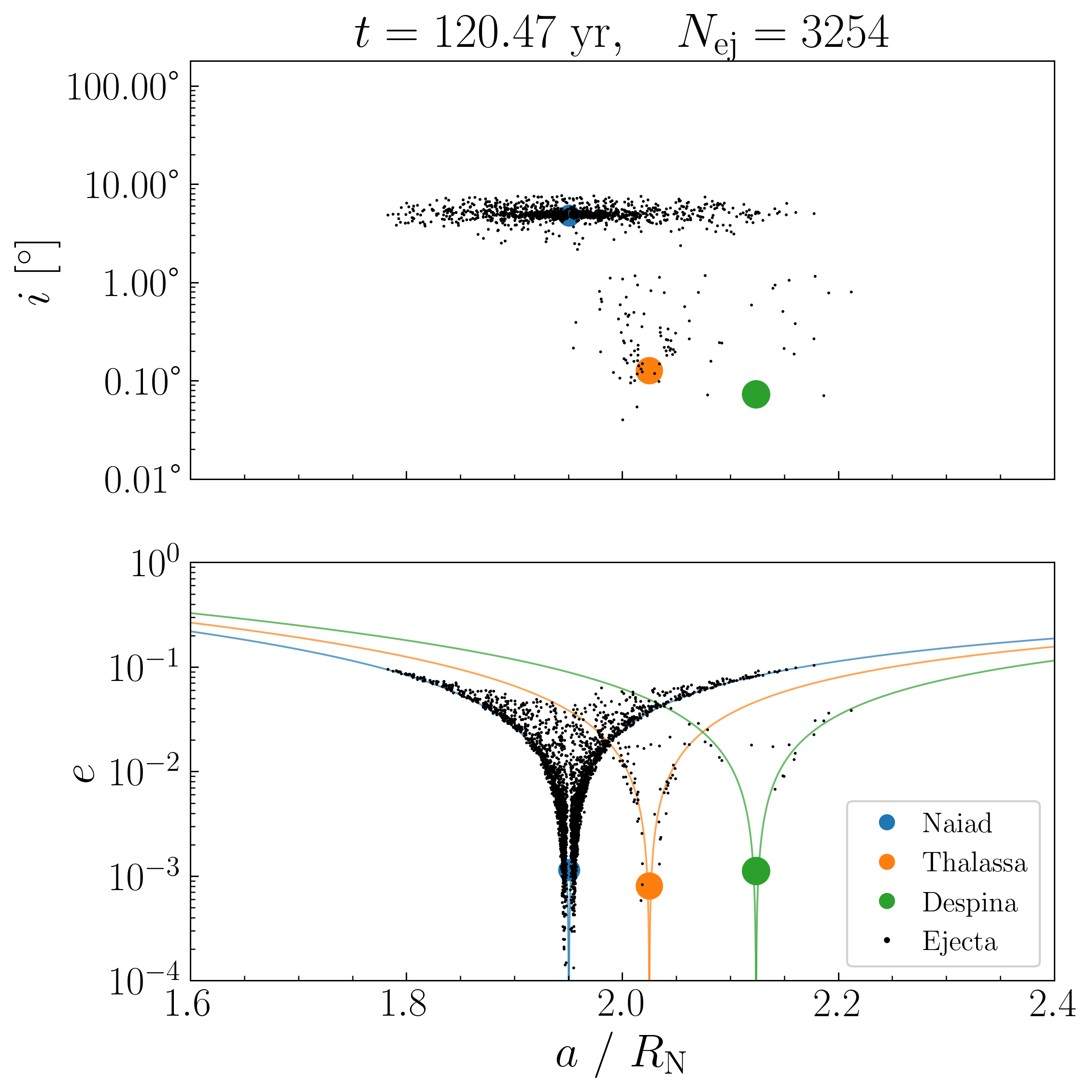}
  \caption{$t \sim 120$ yr}
\end{subfigure}
\begin{subfigure}{0.33\textwidth}
  \centering
  \includegraphics[width=\hsize]{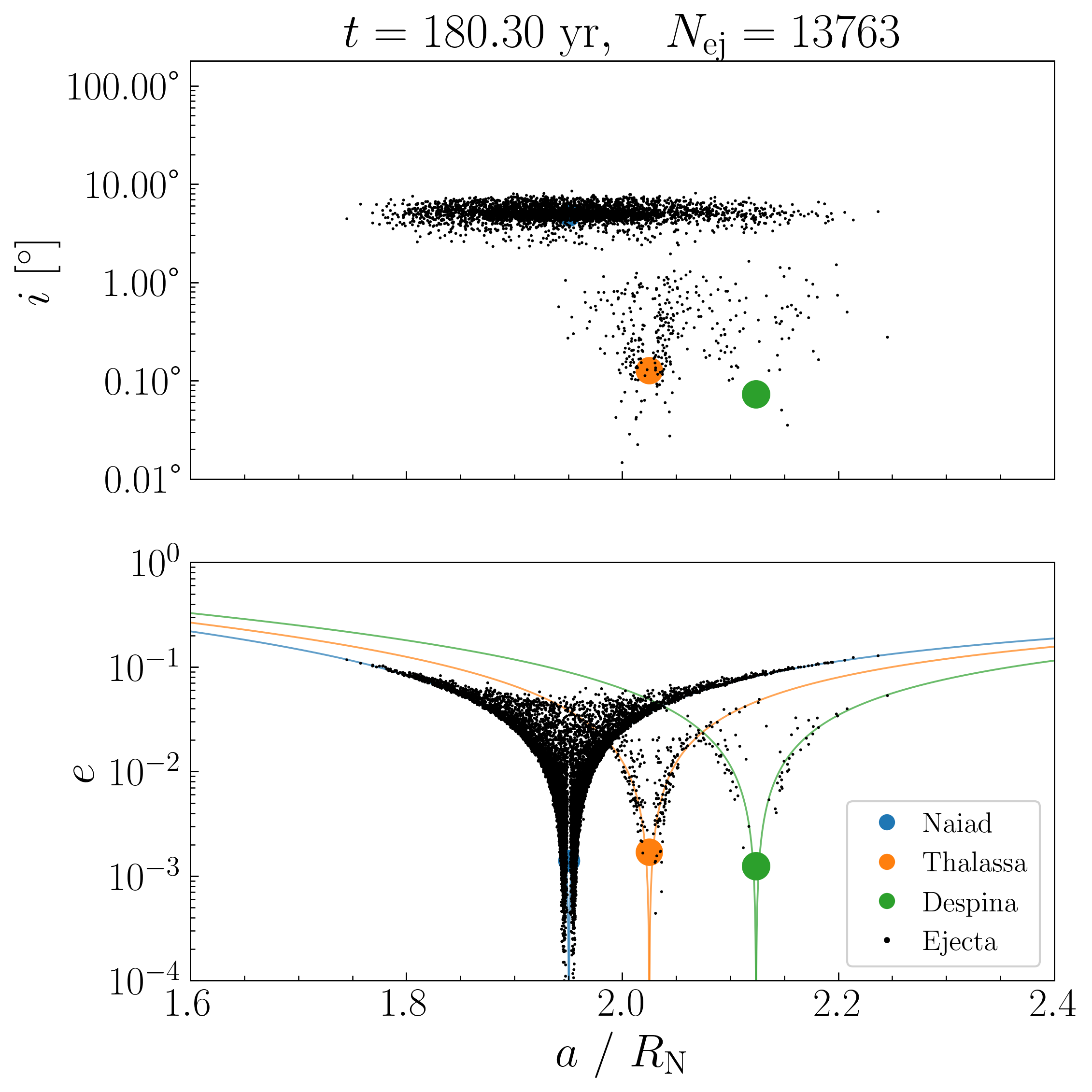}
  \caption{$t \sim 180$ yr }
\end{subfigure}
\caption{\label{fig:snapshots} Snapshots of the orbital element distribution of particles in our simulation with $Y = 1$ MPa. In each plot, the sub-panels show the inclination and eccentricity vs.\ semimajor axis of all particles, where $a$ is given in units of Neptune's radius $R_\text{N}$. Solid curves mark where a particle's pericenter or apocenter equals a given moon's semimajor axis; together they form an envelope enclosing the region of $(a, e)$ space from which a particle could collide with that moon. Panel A shows the system immediately after the heliocentric impact, while panels B–D show the system ${\sim}60$, $120$, and $180$ years later. The number of ejecta particles grows rapidly as Naiad is continuously eroded by collisions. Some ejecta particles reach high enough eccentricities to cross the orbits of Thalassa and Despina and collide with those moons.}
\end{figure*}

In Fig.\ \ref{fig:snapshots}, we show snapshots of the orbital elements $a, e,$ and $i$ of the three moons and all ejecta particles for a simulation with a collision strength of $Y=1$ MPa. At $t=0$, there are ${\sim}200$ ejecta particles resulting from the initial 10 m impactor. In a relatively short amount of time, we see the number of ejecta particles grow rapidly: after 60, 120, and 180 years, the number of ejecta particles has grown to ${\sim}700$, ${\sim}3000$, and ${\sim}14000$, respectively. Even with a strength of 1 MPa, which is extremely high for a small moon thought to be a reaccumulated rubble pile, it seems that Naiad would easily be destroyed by runaway collisional erosion triggered by a small impact.

\begin{figure}
\centering
\includegraphics[width=0.5\textwidth]{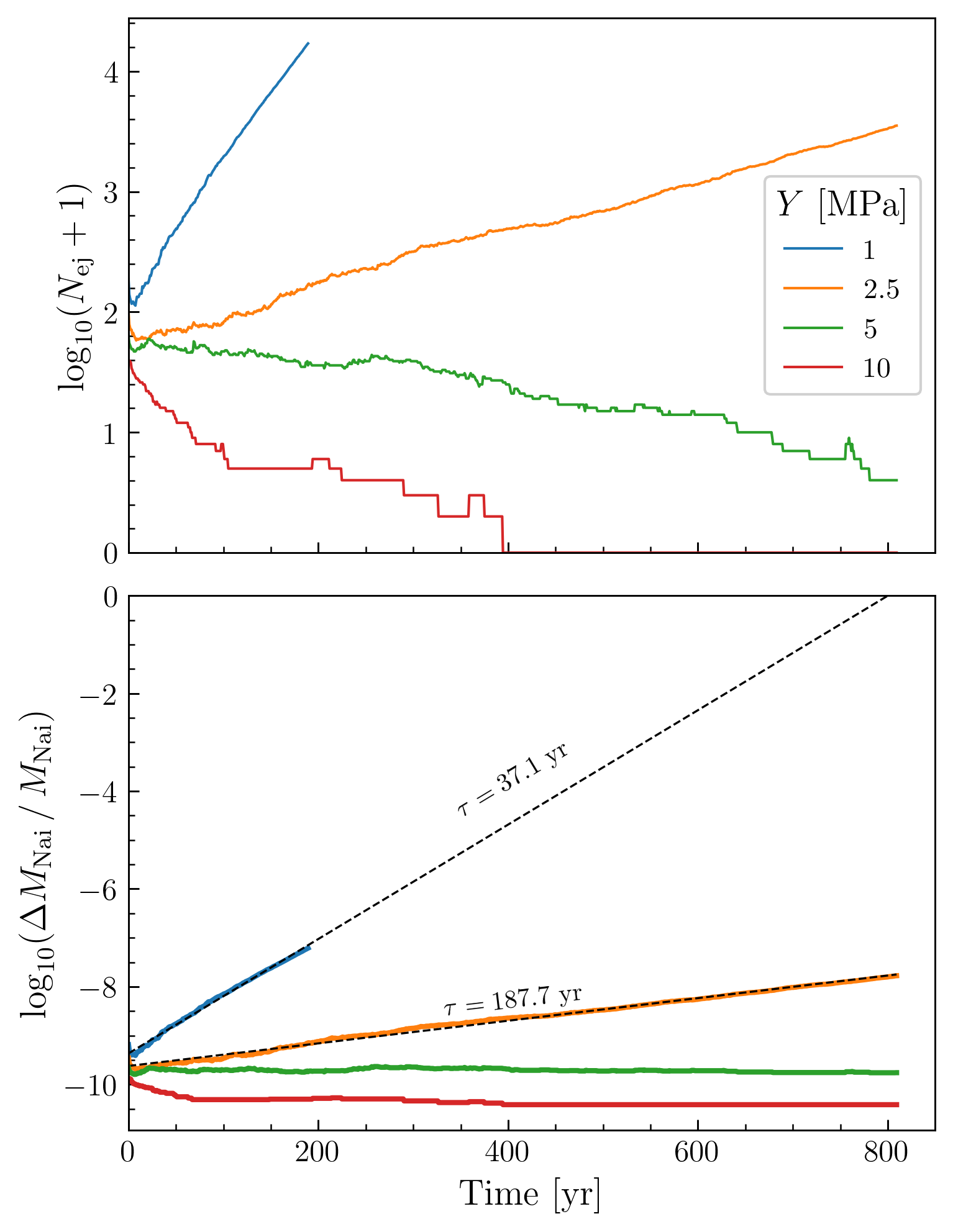}
\caption{\label{fig:timeseries}The time evolution of the number of ejecta particles and the change in Naiad's mass after $10^6$ Naiad orbital periods. The case with 1 MPa of impact strength was halted after ${\sim}$ 200 yr as it became prohibitively slow. }
\end{figure}

To explore the threshold strength to protect Naiad from sesquinary catastrophe, we reran the same simulation varying the collision strength between 1 and 10 MPa. The results are shown in Fig.\ \ref{fig:timeseries}, where we show the number of ejecta particles and the change in Naiad's mass over time. We plot the fractional mass loss as $-\Delta M/M_0 = -(M-M_0)/M_0$ on a logarithmic axis, where exponential growth appears as a straight line with slope $(\tau \ln 10)^{-1}$, with $\tau$ the e-folding timescale. Fig.\ \ref{fig:timeseries} clearly shows exponential mass loss for $Y = 1$ MPa and $Y = 2.5$ MPa. For these two cases, we estimate $\tau$ with a simple linear regression, finding values on the order of decades to centuries. The simulation with $Y=1$ MPa was halted after ${\sim}200$ yr, as the rapid growth in ejecta particles slowed the simulation down to a crawl. However, the short $e$-folding timescale based on the first 200 years demonstrates how rapidly runaway collisional erosion can occur given Naiad's ${\sim}7$ h orbital period. These simulations indicate that Naiad requires a strength of at least several MPa to prevent runaway erosion, implying Naiad's survival to the present day may not be compatible with it being a rubble pile.

\section{Discussion}\label{sec:discussion}

If Neptune's inner moons are the reaccumulated debris from a primordial satellite system disrupted during Triton's capture, then we would naively expect Naiad to be a low-density, rubble pile-like object made of rock and ice with negligible internal strength, possibly similar to the inner moons of Uranus or Saturn. At Naiad's nominal ${\sim}0.8\text{ g cm}^{-3}$, however, it is not gravitationally bound given its elongated shape and requires a cohesive strength on the order of $10^4$ Pa to remain intact. This problem can be alleviated with Naiad's density increased to ${\gtrsim}1.3\text{ g cm}^{-3}$, which would also imply that it is a relatively ice-poor object. Interestingly, this could be consistent with recent spectroscopic observations of Larissa, Galatea, and Proteus showing strong signatures of phyllosilicates and no evidence for water ice \citep{DavisR2026}, raising the intriguing possibility that Neptune's inner moons are not icy at all. Instead, they could be collisional remnants of previous icy satellites that had extensive aqueous alteration in their interiors, and any water ice would have been presumably lost during collisional processing and reaccumulation into the present moons. Future observations that can constrain the compositions as well as sizes (and therefore densities) of Naiad and Thalassa may shed more light on this question.

Even if the ``Roche limit problem'' is resolved, Naiad's contradictions don't end there. Naiad is particularly sensitive to destruction via collisions on timescales that are much shorter than its ${\gtrsim}$ Gyr age implies. It is possible that heliocentric bombardment by SDOs and Centaurs could destroy Naiad on timescales less than 1 Gyr, unless Naiad has enough collisional strength to survive or if the size-frequency distribution of impactors becomes extremely shallow at small sizes. However, given the large uncertainties in the size-frequency distribution of small TNOs and in what Naiad's $Q^*_D$ may be, it is hard to say much about Naiad's survival from heliocentric bombardment. Naiad's bigger problem lies in destruction by planetocentric debris. Our simple $N$-body simulations demonstrated that even small cratering impacts -- which should happen frequently -- can trigger runaway collisional erosion capable of destroying Naiad on decadal timescales, which is incompatible with Naiad's existence today. Naiad's survival therefore requires substantial collisional strength of at least several MPa. Such a high strength is difficult to reconcile with being a reaccumulated rubble pile. One interpretation is that Naiad is a largely coherent, monolithic fragment rather than a gravitational aggregate. This could simultaneously explain its resistance to sesquinary erosion and its survival within the Roche limit. However, it is not obvious how Naiad would survive as a monolithic fragment in the aftermath of the violent disruption of Neptune's primordial satellite system.

There are other possible explanations for Naiad's survival worth considering. First, ejecta–ejecta collisions could grind planetocentric debris down to dust, which could be removed by Poynting-Robertson drag before it has a chance to reimpact Naiad. Second, if Naiad is blanketed by a deep regolith of fine-grained material, small impacts would excavate mostly dust rather than coherent fragments, ensuring that future collisions are not disruptive, although this would require regolith production to outpace regolith removal. In these alternative scenarios, we may expect to see a dusty ring or torus around Naiad unless other mechanisms (i.e., Poynting-Robertson drag) can quickly remove the dust. A detailed assessment of alternative scenarios would require more sophisticated modeling than we attempt here. 

\section{Conclusions} \label{sec:conclusion}

Taken together, these results paint a contradictory picture of Naiad. The same formation scenario that predicts Naiad's existence — reaccumulation of rocky and icy debris from Neptune's primordial satellite system — also predicts physical properties that are difficult to reconcile with Naiad's long-term survival. Either Naiad is physically unusual among small moons, or our understanding of Neptune's inner satellite system is missing something, or both. Among planetary ring-moons, Naiad is arguably one of the most dynamically interesting and is, by several measures, a moon that perhaps should not exist. Resolving this tension will require better observational constraints on Naiad's mass, shape, and composition, and more detailed theoretical modeling of Naiad's dynamical and collisional evolution. A better characterization of Naiad may shed light on the early history of the Neptune system, including the nature of both its present-day and primordial satellites.

\begin{acknowledgements}
We thank Raphael Marschall, Matthew Hedman, Matthew Belyakov, and Ryleigh Davis for useful discussions. H.A. was supported by the French government, through the UCA J.E.D.I. Investments in the Future project managed by the National Research Agency (ANR) with the reference number ANR-15-IDEX-01. H.A. also acknowledges financial support from the Centre national d'\'etudes spatiales (CNES), France (ROR: https://ror.org/04h1h0y33) within the framework of the Hera mission. This work was funded by NASA Emerging Worlds Program award 80NSSC23K1266 (M.C., D.M.) and NASA Merging Worlds grant 80NSSC23K0250 (D.N.). 


\end{acknowledgements}

\bibliographystyle{aa}
\bibliography{references}

\end{document}